\documentstyle[aas2pp4]{article}

\begin{document}

\title{TIDAL DISRUPTION FLARES: THE ACCRETION DISK PHASE}

\author{Matias Montesinos Armijo and Jos\'e A. de Freitas Pacheco}
\author{\it Universit\'e de Nice Sophia-Antipolis - Observatoire de la C\^ote d'Azur\\
Bd de l'Observatoire, BP 4229, 06304 Nice Cedex 4, France }

\begin{abstract}

The evolution of an accretion disk, formed as a consequence of the disruption of a star by a black hole, is followed
by solving numerically hydrodynamic equations. The present investigation aims to study the dependence of resulting
light curves on dynamical and physical properties of such a transient disk during its existence. One of the  main results derived
from our simulations is that blackbody fits of X-ray data tend to overestimate the true mean disk temperature. In fact,
the temperature derived from blackbody fits should be identified with the color X-ray temperature rather than
the average value derived from the true temperature distribution along the disk. The time interval between the beginning of
the circularization of the bound debris and the beginning of the accretion process by the black hole is
determined by the viscous (or accretion) timescale, which also fixes  the rising part of the resulting light curve. 
The luminosity peak coincides with the beginning of matter accretion 
by the black hole and the late evolution of the light curve depends on the evolution of the debris fallback rate.
Peak bolometric luminosities are in the range $10^{45}-10^{46}~\rm erg ~ s^{-1}$ whereas peak luminosities in soft
X-rays (0.2-2.0 keV) are typically one order of magnitude lower. The typical timescale derived from our preferred
models for the flare luminosity to decay by two orders of magnitude is about 3-4 years. Predicted soft X-ray
light curves reproduce quite well data on galaxies in which a variable X-ray emission, possibly related to a tidal
event was detected. In the case of NGC 3599 and IC 3599, data are well reproduced
by models defined by a black hole with mass $\sim 10^7~M_{\odot}$ and a disrupted star of about
one solar mass. The X-ray variation observed in XMMSL1 is consistent with a model defined by a 
black hole with mass $\sim 3\times 10^6~M_{\odot}$ and a disrupted star of one solar mass, while that observed in
the galaxy situated in the cluster A1689 is consistent with a model including a black hole of $\sim 10^7~M_{\odot}$
and a disrupted star of $\sim 0.5~M_{\odot}$.

\end{abstract}

\twocolumn

\section{INTRODUCTION}

Several flare events observed in surveys either in X-rays by {\it ROSAT} (Bade et al. 1996; Komossa
\& Bade 1999), {\it Chandra} and {\it XMM-Newton} (Donley et al. 2002; Esquej et al. 2007, 2008;
Cappelluti et al. 2009, Maksym et al. 2010) or in the UV by {\it GALEX} (Gezari et al. 2006, 2009)
are presently supposed to be associated with the tidal disruption of a star that has passed close to a supermassive
black hole (SMBH), living in a ``dormant" state in the core of the host galaxy. With the advent of several
optical transient surveys such as {\it Pan-STARRS} (Chambers 2007), and the {\it Palomar Transient Factory} (Law
et al. 2009), tidal flare candidates in the optical domain are also expected to be 
discovered. In fact, two possible candidates
have  recently been found in archival data of the Sloan Digital Sky Survey (SDSS; van Velzen et al. 2010). 
Flare properties are linked to the black hole mass and spin, making it an important tool for studying these objects, which otherwise
undetectable due to  the absence of any activity in their ``quiet" phase.
For SMBHs with masses lower than $\sim 10^8~M_{\odot}$, the tidal radius lies outside the horizon of a
Schwarzschild black hole;  this defines the upper limit of masses that can be probed from the analysis of the 
flare light curve.

Studies of the disruption of a star by the tidal field of an SMBH have been made since the late
seventies (Frank 1978; Lacy, Townes \& Hollenback 1982; Nolthenius \& Katz 1982; Bicknell \& Gingold 1983;
Carter \& Luminet 1983; Rees 1988; Ulmer 1999; Ayal, Livio \& Piran 2000) and a general picture describing
the sequence of events has now emerged. The physics of the disruption process is essentially described in terms
of two dimensionless parameters: the ``penetration" parameter $\lambda=R_t/R_p$, defined by the ratio 
between the tidal radius $R_t$ and the periapse distance $R_p$ and the parameter $\eta$ introduced by 
Press \& Teukolsky (1977), defined by the square root of the ratio of the self gravity of
the star to the tidal gravity at the surface when the star passes by the periapse.  
If $\lambda \geq 5$, during the periapse passage the star experiences a strong and rapid compression along directions
perpendicular to the orbital plane and azimuthally, while stretching in the radial 
direction, leading to an increase of the central density by a factor $\sim \lambda^3$ (Carter \& Luminet 1983).
Using a ``smoothed particle hydrodynamics" (SPH) code, Nolthenius \& Katz (1982) investigated the tidal effects
on a 1.0 $M_{\odot}$ star, modeled by a polytrope of index $n=1.5$, in a parabolic orbit around 
a $10^4~M_{\odot}$ black hole. According to their simulations, the star becomes marginally disrupted
if $\eta \geq 1.6-5.8$. A similar analysis was performed by Khokhlov, Novikov \& Pethick (1993), who 
used a three-dimensional Eulerian hydrodynamical code. For a star described by a polytrope of index $n=1.5$, they found that
the dissipated tidal energy is comparable to the unperturbed binding energy if $\eta \sim 1.0-1.5$. For
higher values of $\eta$ the star is completely disrupted and otherwise  partially stripped.

Early investigations developed a scenario in which the sudden tidal compression could lead to an increase in the thermonuclear 
energy generation rate,  consequently increasing the luminosity of the star and 
eventually producing a supernova-like explosion (Carter \& Luminet 1983; Pichon 1985). However, subsequent numerical simulations 
have shown that the attained compression ratio is much smaller and that the triple-$\alpha$ reaction
is not ignited, excluding a possible detonation of the star (Bicknell \& Gingold 1983). However, for a periapse passage
with $\lambda\sim 35$, significant processing of C,N,O can occur and the energy released by the thermonuclear reactions
involving these elements could be comparable to the binding energy of the star, raising the possibility that the
gas in the vicinity of the central BH may have a non-solar abundance pattern (Bicknell \& Gingold 1983). 

For a near parabolic orbit leading to a close encounter and to a tidal disruption of the star, there are two
distinct phases in the dynamical evolution of the debris, in which an important amount of energy is released.
The first is the ``fallback" phase. The kinetic energy of the
debris is much more important than its self-gravity and its internal energy. Nearly half of the debris
is bound,  initially forming gas streams with approximately ballistic trajectories, having a large spread in orbital
periods and a mass distribution as a function of the energy nearly constant (Evans \& Kochanek 1989; Ayal, Livio
\& Piran 2000). The most bound material returns to pericenter after a timescale $t_{\rm min}$ which, in 
the majority of the cases of interest, is of the order of few years. The return of the bound debris initiates the formation 
of a small accretion disk around the BH. The circularization of the bound material is expected to be efficient 
since it  is mainly driven  by shocks between gas streams that converge at 
periapse (Evans \& Kochanek 1989; Laguna et al. 1993). The total energy radiated in this phase is of the 
order of $GM_{\rm bh}\Delta m_*/2R_t$,
where $\Delta m_*$ is the mass of the bound material. For a BH of mass of $10^7~M_{\odot}$ and $\Delta m_* \sim$ 0.5 
$M_{\odot}$, this corresponds to an energy of about $2\times 10^{52}$ erg that, as we shall see later, represents 
about 10\% of the total energy released in the later evolutionary period.
The second, or ``viscous",  phase initiates as soon as the disk begins to  form, since
the viscous forces responsible for the process of angular momentum transfer throughout the disk dissipate kinetic 
energy. By this mechanism the gas is heated and radiates; this is probably the major source of the flare luminosity. 

The rate at which the bound debris returns to periapse and circularizes is often assumed 
to be identical to the rate at which the disk is formed or to the rate at which the BH accretes
mass, radiating close or above the Eddington limit (Ulmer 1999). 
In fact, we will show that there is a time delay between the beginning of circularization and the
instant the BH begins to accrete mass. During this short initial accretion phase, the BH accretes mass at
a rate considerably higher than the fallback rate, and only in the final evolutionary
stages are  the rates of both processes  comparable. 
A more detailed analysis of the disk properties was performed by Cannizzo, Lee \& Goodman (1990), who 
solved a diffusion-like equation describing the evolution of the surface density. They concluded  that
the disk remains very luminous for several thousand years, representing a non negligible fraction of the
time between successive stellar disruptions. Similar calculations were performed by Strubbe \& Quataert (2009), who 
have also included in their model the reprocessing of the UV radiation by the unbound debris. As a consequence
of such a process, a rich spectrum of very broad emission lines is produced.

More recently, Montesinos \& de Freitas Pacheco (2011, hereafter MP11) studied the evolution of self gravitating
disks around SMBHs, solving numerically the complete set of hydrodynamical equations. In the present paper, we report
an investigation of the evolution of the accretion disk formed just after a tidal disruption event, using the code
described in MP11. We will show that once the circularization process begins, two evolutionary phases can be distinguished.
The first is an early and short phase in which the black hole is not accreting mass since the debris has not yet reached the last
stable circular orbit, which occurs in a time fixed by the accretion or viscous timescale. 
In this phase, the flare begins to develop with the luminosity of the forming disk rising very rapidly.  
The second phase initiates once the black hole begins to accrete mass, which coincides with the instant
of maximum flare luminosity. In its late evolution, the disk luminosity decays approximately at the fallback rate.
The flare (and the disk) duration in the present models is typically of the order of a few up to tens of years, depending
on the fallback rate. These values are considerably 
shorter than the timescales derived by Cannizzo, Lee \& Goodman (1990). The paper is organized as 
follows: In Section 2 we discuss  the model
and equations;  in Section 3 we present  the main results and, finally, in Section 4 we give our conclusions.

\section{THE DISK MODEL}

Past investigations of  the evolution of the accretion disk phase described the flow and the energy
transfer by using several approximations. Moreover, most of these studies assumed that the disk is 
practically in place when its evolution begins. In fact, the accretion timescale
$t_a\sim r/V_r$ differs from the fallback timescale of the bound debris, implying that there is
a time lag between the beginning of the circularization of the bound material and the beginning of the accretion process
by the central black hole. When the accretion process by the black hole begins, most of the bound material 
is not yet settled in orbit and, as we shall see, only 
in the late phases does the SMBH accrete mass at a rate comparable to the fallback rate.
 
The set of hydrodynamical equations (in cylindrical coordinates) describing the disk evolution as well as
the details of the numerical code can be found in MP11 and here only some general aspects of the code
are reviewed. The code is based on an eulerian formalism, using a finite difference method of second-order
according to the Van Leer upwind algorithm on a staggered mesh. This means that scalar quantities such
as surface density and scale of height are defined at the center of a cell, whereas vector quantities such as
velocity and fluxes are defined at the interface between cells. Since disks considered in the present
study are considerably smaller than those investigated by MP11, which have dimensions of $\sim$ 50 pc,
we have adopted an integration grid (one-dimensional) with 256 ring sectors instead of the original 1024 rings adopted 
by MP11. The inner radius of the grid coincides with the last stable circular orbit ($r_{\rm lso}=6GM_{\rm bh}/c^2$)
while the external radius is defined by the tidal radius. The time step is controlled by the Courant-Friedrich-Levy
(CFL) condition, which states that the information cannot sweep a distance greater than the size of a
cell. For further details, see MP11. Note that our axisymmetric geometry does not permit us to follow
the debris evolution before circularization, when streams have eccentric orbits. 

Other particular aspects should be mentioned. The first refers to the modeling of the viscosity responsible 
for the angular momentum transport, which
in all past investigations was assumed to be described by the so-called ``$\alpha$-model" introduced
by Shakura \& Sunyaev (1973). In this approach, the kinematic viscosity $\eta$ (hereafter $\eta$ represents
the kinematic viscosity and not the Press-Teukolsky parameter, unless stated explicitly) due to the subsonic
turbulence is given by
\begin{equation}
\eta = \alpha Hc_s
\end{equation}
where $\alpha \leq 1$ is a dimensionless coefficient, $H$ is the vertical scale of the disk, supposed to be
of the same order as the typical (isotropic) turbulence scale $\ell_t$ and $c_s$ is the sound velocity.
As discussed in MP11, if the flow is self-regulated, it must be characterized by a critical Reynolds number
${\cal R}$ determined by the viscosity above which the flow becomes unstable. Under these conditions,
using the formalism developed by de Freitas Pacheco \& Steiner (1976), the kinematic viscosity can be
parametrized  as
\begin{equation}
\eta = \frac{2\pi r V_{\phi}}{{\cal R}}
\end{equation}
where $r$ is the radial distance to the center of the disk and $V_{\phi}$ is the azimuthal velocity of
the flow at that position. According to the analysis by Piran (1978), accretion disks modeled by such a viscosity
prescription are thermally stable, while this is not the case for $\alpha$-model disks.

The second aspect concerns the gravitational potential in the disk that, in the present case, is due
essentially to the central BH since the disk self gravity is negligible. As in MP11, it was assumed
here that the gravitational field of the BH is represented by the approximate potential of Paczynski-Wiita 
(Paczynski \& Wiita 1980).

The third point concerns the modelization of the matter flux due to the fallback of the bound material. 
As mentioned above, the disk formed by the debris is supposed to have an extension comparable to
the tidal radius, where we have assumed that an inward flux of matter is present. 
The timescale of the circularization process is still quite uncertain although some authors consider
that it should be comparable to $t_{\rm min}$ (Ulmer 1999). As we have mentioned previously, $t_{\rm min}$ is
of the order of a few years while the dynamical timescale at the level of the tidal radius is 
of the order of a day or even less (note that if we had adopted $\mu$=1 in Equation ~\ref{tidal} below,
the resulting timescale would be less than one year). Thus, we cannot exclude the possibility that the debris material
circularizes in a timescale shorter than the accretion timescale and, in fact, here 
we have assumed that this occurs. Under these conditions,
a modification in the outer boundary condition was introduced with respect to those discussed in MP11.
In that work the boundary conditions were fixed by the following considerations: the (inward) mass flux
at the inner radius (last stable orbit) corresponds to the mass accreted by the black hole and lost
by the disk while at the outer radius the inward flux is taken to be zero, i.e., there is no
source of matter for the disk. This is not the case in the present investigation since there is
a continuous and variable source of matter due to the circularization process. Consequently, the original
external boundary condition was modified and an inflow is now allowed 
such that the total rate of matter flowing into the disk is equal to the total rate at which
the debris returns to periapse. According to early investigations by Rees (1988) and Evans \& Kochanek (1989)
the fallback rate is given by the relation 
\begin{equation}
\label{debrisrate1}
\frac{dM}{dt}\simeq \frac{1}{3}\frac{m_*}{t_{\rm min}}\left(\frac{t_{\rm min}}{t}\right)^{5/3}
\end{equation}
where $m_*$ is the mass of the disrupted star and
\begin{equation}
\label{tmin}
t_{\rm min}= \frac{\pi}{\sqrt{2}}\frac{R_p^3}{(GM_{\rm bh}R_*^3)^{1/2}}
\end{equation}
with $R_*$ being the radius of the unperturbed star configuration. Here the periapse distance will be
taken to be equal to the tidal radius $R_t$, which was computed from the relation
\begin{equation}
\label{tidal}
R_t=\mu R_*\left(\frac{M_{\rm bh}}{m_*}\right)^{1/3}
\end{equation}
Most of authors in the literature take $\mu$=1 but here we will take $\mu\simeq$ 2.4, corresponding approximately to the
Roche model. 

A more detailed analysis by Lodato, King \& Pringle (2009) indicates deviations from Equation ~\ref{debrisrate1} 
at the early phases. According to them, only in the late evolutionary stages does the fallback rate 
vary as $t^{-5/3}$. As an attempt
to include such deviations, we have adopted in our computations the following formula for the fallback rate
\begin{equation}
\label{debrisrate2}
\frac{dM}{dt} = A\frac{(t/t_{\rm min})^{1/3}}{\left[a+(t/t_{\rm min})^2\right]}
\end{equation}
where $A$ is a normalization constant and $t_{\rm min}$ is still given by 
Equation ~\ref{tmin}. Imposing that the time integral of Equation ~\ref{debrisrate2} be equal to $m_*/2$, which is the
same condition satisfied by Equation ~\ref{debrisrate1}, one obtains $A=\sqrt{3}a^{1/3}m_*/(2\pi t_{\rm min})$.  
The dimensionless parameter $a$ permits control of  the instant at which the maximum fallback rate
occurs since this is given by $t_{\rm max}= \sqrt{a/5}t_{\rm min}$. Note that the late decay given by Equation ~\ref{debrisrate2}
is proportional to $t^{-5/3}$ as in Equation ~\ref{debrisrate1}. Equation ~\ref{debrisrate2} has also some 
computational advantages since the beginning of the circularization process can be taken at $t=0$.

For technical reasons, we assume that  the BH is initially  surrounded by a very faint ``phantom" disk, having
a negligible mass in comparison with that of the bound debris. In fact, such a putative ``phantom" disk is 
introduced only to avoid initially the presence of zeros in the integration grid; this does  not affect 
the final results. 
 
Two series of models were computed: those labeled A for which $a=0.01$ and those labeled M
for which $a=1$. For M-models, the black hole always  begins to accrete  before the occurrence of the
maximum of the fallback rate, while for A-models, depending on
the adopted critical Reynolds number, the black hole begins to accrete close to or after the occurrence of the
maximum of the fallback rate. As we shall see below, the flare light curves derived from A  and M
series differ considerably, giving   important information about the accretion timescale.
Each computed disk model is characterized by the critical Reynolds number, the SMBH mass, the 
mass, and the radius of the disrupted star. 
These parameters for different computed models are given in Table 1.

\begin{table}
\caption{Model parameters: the first column identifies the model, the second gives
the black hole mass, the third and the fourth give respectively the mass and radius
of the disrupted star while the fifth gives the critical Reynolds number.}             
\centering                         
\begin{tabular}{c c c c c}        
\hline\hline                 
Model&$M_{bh}/M_{\odot}$&$m_*/M_{\odot}$&$R_*/R_{\odot}$&${\cal R}$\\ 
\hline                        
A1, M1 & $1\times 10^7$ & 1.0 &1.0& 500\\     
A2, M2 & $1\times 10^7$ & 1.0 &1.0&1000\\
A3, M3 & $1\times 10^7$ & 1.0 &1.0&1500\\
A4, M4 & $1\times 10^7$ & 2.0 & 1.7&1000\\
  A5   & $1\times 10^6$ & 1.0 & 1.0&500\\
A6, M6 & $3\times 10^6$ & 1.0 & 1.0&500\\
A7, M7 & $3\times 10^6$ & 2.0 & 1.7&500\\
A8, M8 & $3\times 10^6$ & 2.0 & 1.7&1000\\
  A9   & $3\times 10^6$ & 1.0 & 1.0&1000\\
 A10   & $3\times 10^6$ & 1.0 & 1.0& 100\\
 A11   & $1\times 10^7$ & 0.5 & 0.6& 100\\
 A12   & $1\times 10^7$ & 0.5 & 0.6&1000\\
\hline                                   
\end{tabular}
\end{table}

\section{RESULTS}

\subsection{The Disk Formation}

Once the bound material begins to circularize (this corresponds to $t=0$, the beginning of our integration procedure), the 
disk develops gradually and is formed in a time interval
controlled essentially by the accretion timescale $r/V_r$, where $r$ is equal to the tidal radius.

 This gradual formation of the disk 
is illustrated in Figure ~1, where several snapshots of the surface density 
profile derived from model A2 are shown. In the early evolutionary phases, the forming disk is 
characterized by a ``torus-like" structure whose effective internal radius decreases as the material inspirals towards
the black hole. The surface density of this structure increases in time as the 
front approaches the last stable  orbit; This is a consequence of the fact that in these initial phases matter is 
deposited in the outer region of the disk at a rate higher than that which is conveyed inward. After reaching the last
stable orbit (density profile labeled as ``5" in Figure ~1), the disk becomes more or less homogeneous and
its late evolution is characterized by a decreasing  surface density, which is a  consequence of the exhaustion of the
debris reservoir.

\begin{figure}
\plotone{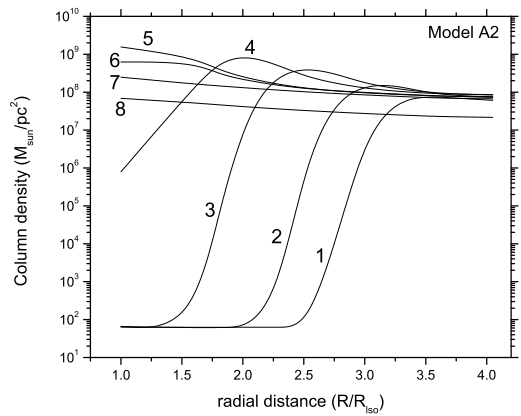}
\caption{Different snapshots showing the evolution of the surface density. Distances are
given in units of the last stable circular orbit. Labels correspond
to the following instants of time: 1=0.0455 yr; 2=0.091 yr; 3=0.1823 yr; 4=0.273 yr; 5=0.4102 yr; 6=0.4558 yr;
7=0.501 yr; 8=1.367 yr. Profile 5 corresponds to the beginning of the accretion process by the black hole.}
\end{figure}

During the initial phases when the disk is still being formed and the black hole is not 
accreting mass, the flow of matter throughout the disk is not uniform. This can be seen in Figure~2 where 
snapshots of the mass flow profile  for model A2 are also shown. The flow in the inner regions increases in amplitude
as it propagates inwards and, when it reaches the last stable orbit (profile ``5" in Figure~2), the accretion
process by the black hole begins and a short, but significant peak in the luminosity curve is produced as we
shall see below. Late in the process, the flow becomes practically uniform 
but with a decreasing amplitude, following the fallback rate of decay. 

\begin{figure}
\plotone{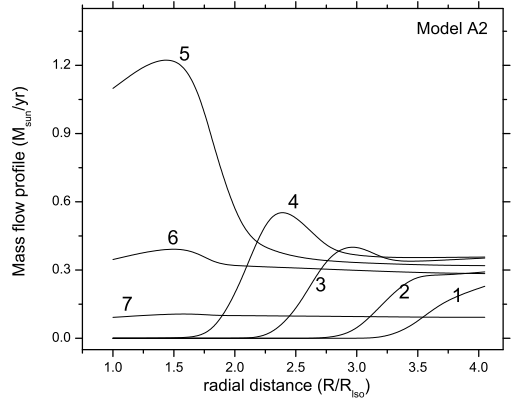}
\caption{Snapshots showing the evolution of the mass flow. Distances are given in units of the
last stable circular orbit. Labels correspond
to the following instants of time: 1=0.0455 yr; 2=0.091 yr; 3=0.1823 yr; 4=0.273 yr; 5=0.4102 yr; 6=0.4558 yr;
7=0.501 yr. Profile 5 corresponds to the beginning of the accretion process by the black hole.}
\end{figure}

In Figure~3 the evolution of the accretion rate by the central black hole is plotted for models A1, A2, A3 and, for comparison, 
that of the fallback rate. Note  that when the debris reaches the last stable
orbit and the black hole begins to accrete mass, there is a sudden peak in the accretion rate which surpasses 
the fallback rate at that instant. Such
a peak is in fact expected since its time integral corresponds approximately to the debris mass deposited up to
that instant inside the disk. After the peak maximum, the accretion rate decays very rapidly and joins smoothly 
the fallback rate curve, initiating
a late phase in which the accretion rate by the black hole is equal to the fallback rate. This corresponds to phases
in which the matter flow through the disk is practically uniform and the luminosity is proportional to the
accretion rate by the BH. Note  that the peak in the accretion rate for 
models A1, A2, and A3 occurs at different times;  the explanation for such a behavior is quite simple. These models 
have the same parameters but have flows characterized by different critical Reynolds numbers. 
Since the radial velocity scales as $V_r \propto V_{\phi}/{\cal R}$, the accretion timescale for these
three models differs only in ${\cal R}$.
Thus, one should expect that the ratio between instants of maximum is equal to the ratio of the corresponding
Reynolds number defining the model. The derived instants of maximum accretion for both series A and M are given in
Table 2. For models A1, A2, and A3 the accretion peak occurs respectively at 0.205 yr, 0.415 yr, and
0.606 yr and their ratios follow quite well the ratios between 500, 1000 and 1500, the critical Reynolds number 
characterizing these models. From table 2, it is easy  verify that these 
proportional ratios also hold for models M1, M2, and M3.

\begin{figure}
\plotone{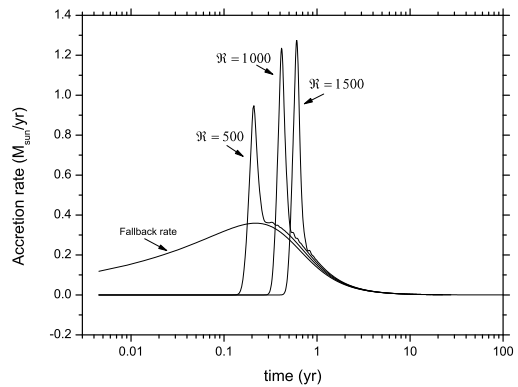}
\caption{The evolution of the accretion rate by the black hole for models differing only in the critical Reynolds
number. The evolution of the fallback rate for $a=0.01$ is also shown.}
\end{figure}

The accretion timescale controlling the disk formation 
 scales as $t_a \propto {\cal R}\left(R_*^{3/2}m_*^{-1/3}\right)$, independent of the 
black hole mass. Models A1, A5, and A6 have the same parameters but differ in the black hole masses. 
As expected, the
instant of maximum accretion derived for these models is nearly the same as that which can be verified by simple
inspection of Table 2. A further check can be performed by plotting the derived instant of maximum
accretion $t_p$ as a function of $t_a$ expressed in  terms of the critical Reynolds number and
parameters of the disrupted star. This was done in Figure ~4 for models of series ``A''.
As expected, the correlation is
highly significant (correlation coefficient 99.6\%), providing a further test for the adequacy of the
algorithm adopted for the numerical solution of the hydrodynamic equations.
The solid line in the plot corresponds to the
relation
\begin{equation}
\label{fitpeak}
\log t_p = -3.381 + \log{\cal R}\frac{R_*^{3/2}}{m_*^{1/2}}
\end{equation}

\begin{figure}
\plotone{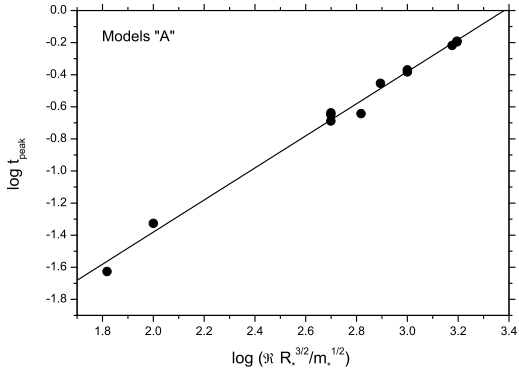}
\caption{Instant of the peak in the accretion rate versus the accretion timescale. Filled dots correspond
to values derived from ``A" models and solid line to the relation given in the text.}
\end{figure}

\begin{table}
\caption{Disk and flare properties: the first column identifies the model, the second gives
the ratio between the tidal and the last stable orbit radii, a measure of the disk extension. The third
column gives $t_{min}$ in years
and the last two columns give the instant of the maximum accretion in years respectively for
models of series $``A"$ and $``M"$.}             
\centering                         
\begin{tabular}{c c c c c}        
\hline\hline                 
Model&$R_t/R_{lso}$&$t_{min}$&$t_p (A)$&$t_p(M)$\\ 
\hline                        
   1 & 4.05 & 4.905 & 0.205 & 0.228\\     
   2 & 4.05 & 4.905 & 0.415 & 0.474\\
   3 & 4.05 & 4.905 & 0.606 & 0.684\\
   4 & 5.47 & 5.436 & 0.638 & 0.410\\
   5 & 18.84& 1.551 & 0.231 &  - \\
   6 & 9.06 & 2.686 & 0.225 & 0.315\\
   7 & 12.22& 2.977 & 0.352 & 0.498\\
   8 & 12.22& 2.977 & 0.644 & 0.779\\
   9 & 9.06 & 2.686 & 0.427 &  - \\
  10 & 9.06 & 2.686 & 0.047 &  - \\
  11 & 3.07 & 4.559 & 0.024 &  - \\
  12 & 3.07 & 4.559 & 0.228 &  - \\ 
\hline                                   
\end{tabular}
\end{table}

\subsection{The Light Curve}

The shape of the light curve depends strongly on the ratio between the accretion time $t_a$ and the
instant of maximum fallback rate $t_{\rm max}$. If $t_a/t_{\rm max} < 1$, i.e., the material reaches the
last stable orbit before the occurrence of the maximum fallback rate (all M  models satisfy this condition), 
a ``bump" in the light curve appears just after the luminosity peak. This feature can
be explained in the following way: The luminosity maximum coincides with the peak in the accretion rate.
After the peak, the BH accretion rate has a
fast decay and  gradually joins  the fallback rate curve. Since the maximum rate of the fallback material
has not yet occurred, the accretion rate will pass by the secondary maximum, producing the aforementioned ``bump". 
In the opposite case ($t_a/t_{\rm max} > 1$), satisfied by most of the  ``A'' models, such a ``bump"
does not appear because the black hole begins to accrete after the occurrence of the maximum in the fallback rate. In this
case, after the peak, the luminosity decays as $t^{-5/3}$ similar to the late evolution of models ``M''. This 
behavior is shown in Figure ~5 where the bolometric light curves for models A4 and M4 are shown for comparison.
Such a trend differs from that derived from the similarity solution obtained by Cannizzo, Lee \& Goodman (1990), which 
indicates a late luminosity decay as $t^{-19/16}$.

\begin{figure}
\plotone{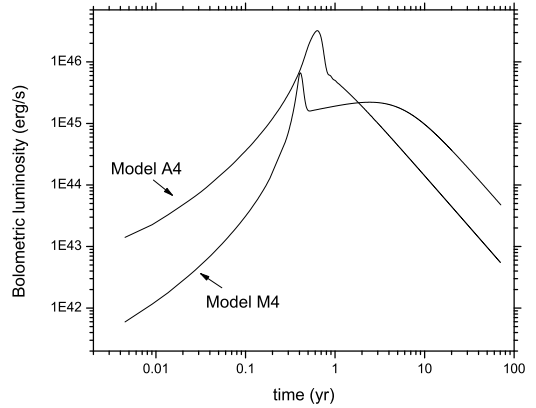}
\caption{Flare light curves for two models having the same parameters but with a different evolution in the
fallback rate. Notice the presence of a ``bump"  just after the peak luminosity in the light curve of model M4.}
\end{figure}

The characteristic timescale for the luminosity to decay by two orders of magnitude after the peak is 
also different for models of series ``A'' and ``M''. The former have on  average typical decay timescales 
of about 3.2 yr, while for the later values are on  average around 43.3 yr. The narrow peak feature is
also shorter for ``A'' models ($\sim$ 5.5 months against $\sim$ 8.3 months on the average for M models).
Moreover,   ``A''  models generally produce  flares with peak luminosities of the order of few $10^{46}~\rm  erg ~ s^{-1}$
while the peak luminosity of models ``M'' is, on average, one order of magnitude lower.
As expected, peak bolometric luminosities $L_p$ correlate quite well
with the peak accretion rate ${\dot R}_p$ by the central black hole. For models ``A", a best fit of 
derived data gives
\begin{equation}
\log L_p = (46.209\pm 0.031)+(1.060\pm 0.110)\log{\dot R}_p \pm 0.106
\end{equation}
with a correlation coefficient of 95.0\%. In this relation the peak luminosity is in $\rm erg ~ s^{-1}$ and
the peak accretion rate is in $M_{\odot} \rm yr^{-1}$. The total energy radiated by the event
is proportional to the total accreted mass by the black hole ($E \sim \Delta m_*c^2$), being of
the order of few $10^{53}$ erg.

It should be emphasized again that during the rising phase of the flare, the black hole is not accreting mass. The
radiation comes from the material heated by viscous dissipation in a disk still in formation and far from 
equilibrium. When the accretion process begins, the resulting luminosity peak exceeds the
Eddington value. However, one may wonder whether the supercritical Eddington regime should be applied to
a disk geometry (see, for instance, Heinzeller \& Duschl 2007). In the case of a disk, the condition for 
having equilibrium along the $z$-axis is local. Near the last stable orbit, tidal
forces due to the SMBH and pressure gradients balance each other and the condition for stability, i.e.,
the condition for radiation pressure gradients to  not overtake gravity is (MP11)
\begin{equation}
Q_{\rm rad}(r) < \frac{cGM_{\rm bh}}{2r^2}\frac{\Sigma}{\tau_{\rm ef}}\left(\frac{H}{r}\right)
\end{equation}
where $Q_{\rm rad}$ is the radiative flux, $\Sigma$ is the surface density, $\tau_{\rm ef}$ is the
effective optical depth and $H$ is the disk scale of height. The present models satisfy this
condition for  the following reason: As the radiation pressure begins to inflate the disk, the 
fraction of advected energy increases, permitting higher accretion rates without increasing the 
radiative flux. It should be emphasized that this cooling effect 
is strengthened by the photon-trapping mechanism, as first pointed out by 
Begelman (1978) and Oshuga et al. (2002).

\subsection{Physical Properties of the Disk}

The evolution of the surface density is similar for all computed models as discussed 
previously. In this section, the evolution of the disk temperature will be examined.
The local effective temperature of the disk was derived by equating the energy rate per unit of
area dissipated by viscous forces, corrected by advection and photon-trapping, to the
energy rate per unit of area radiated away (see MP11 for details). Inside the disk, photons interact 
with matter essentially by scattering on free electrons and by free-free processes. The effective
optical depth of the photon along the z-axis is $\tau_{\rm ef}=\sqrt{3\tau_{\rm ff}(\tau_{\rm ff}+\tau_{\rm s})}$, where $\tau_{\rm ff}$
is the optical depth due to free-free processes and $\tau_{\rm s}$ is the Thomson
scattering optical depth. Figure~6 shows snapshots of the evolution of the effective optical depth profile
for model A2. Note that either the evolving ``torus"-like structure or the disk when formed (profile 
labeled ``5") is always optically thick, but the integrated disk spectrum is not that of 
a pure blackbody for two main reasons: firstly the effective temperature varies across the
disk surface and secondly, on the average, $\tau_s/\tau_{\rm ff} \sim$ 1 throughout the disk. Deviations from
a blackbody spectrum are more severe when $\tau_s/\tau_{\rm ff} >> 1$ since in this case the local flux is
approximately given by $F(\nu) \propto B_{\nu}(T)\sqrt{\tau_{\rm ff}(\nu)/\tau_{\rm s}}$
 
\begin{figure}
\plotone{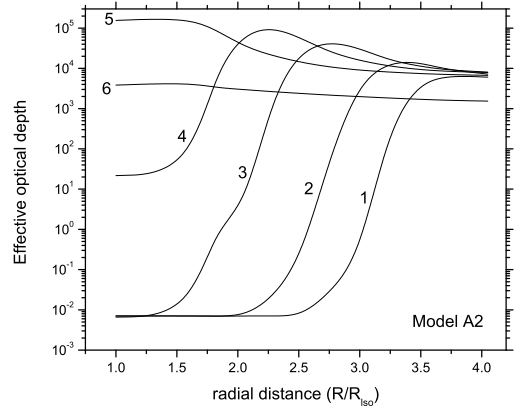}
\caption{Evolution of the effective optical depth profile for model A2. Distances are in units
of the last stable circular orbit. Labels correspond to instants: 1=0.0455 yr; 2=0.091 yr; 3=0.1823 yr;
4=0.273 yr; 5=0.4102 yr; 6=1.367 yr.}
\end{figure}

The evolution of the effective temperature profile for model A2 is shown in Figure ~7. Since the dissipated
energy rate per unit of area due to viscosity is proportional to the surface density, the
evolution of the effective temperature profile follows the former. In the early phases, profiles indicate
that the temperature increases inwards and in time as the material approaches the last stable orbit. The
maximum value is attained in the inner region of the disk at the instant when the black hole begins to accrete mass
(see profile ``5" in Figure ~7). All models show the same trend but the temperature values depend on the adopted
parameter, in particular that of the black hole mass.
  
\begin{figure}
\plotone{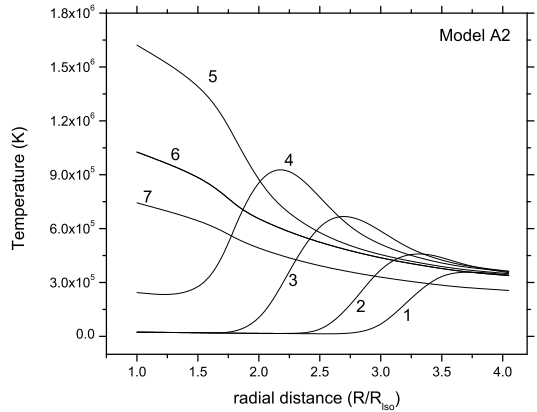}
\caption{Evolution of the effective temperature profile for model A2. Distances are in units
of the last stable circular orbit. Labels correspond to instants: 1=0.0455 yr; 2=0.091 yr; 3=0.1823 yr;
4=0.273 yr; 5=0.4102 yr; 6=0.4558 yr; 7=1.367}
\end{figure}
\begin{figure}
\plotone{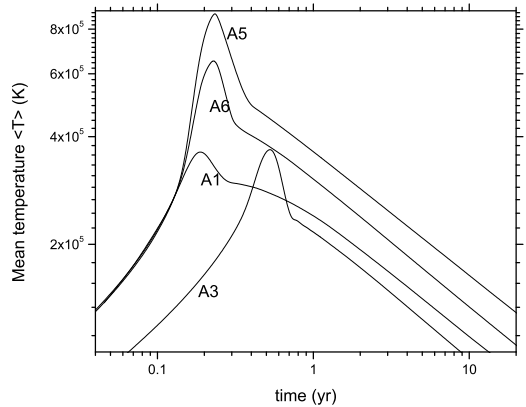}
\caption{Evolution of the average effective temperature of the disk for different models.}
\end{figure}

Observers fit, in general, the spectral data of flare candidates by an equivalent blackbody 
distribution characterized by a unique temperature. As seen in Figure ~7, the effective
temperature in the early evolutionary phases varies considerably along the disk surface and only in the late
evolutionary stages does it become more uniform. In order to compare temperatures derived from our models
with some real data, we have defined a suitable mean effective temperature for the disk  by the relation
$<T> \equiv <T^4>^{1/4}$ where
\begin{equation}
<T^4> = \frac{\int^{R_t}_{R_{\rm lso}}rT^4(r)dr}{\int^{R_t}_{R_{\rm lso}}rdr}
\end{equation}

The evolution of such a mean temperature for models A1, A5, A6 and A3 is shown in Figure ~8. Models
A1, A5 and A6 have the same parameters but differ on the BH mass. Despite the fact that in the first 2
months of the flare evolution, when the disk is still being formed, there is no difference among 
these three models, the maximum attained temperature
and the late evolution differ in the sense that models with more massive black holes produce
``colder" disks, since the dissipation rate due to viscous forces is less important.
In fact, our models indicate that $<T>_{\rm peak} \propto M_{bh}^{-5/2}$. For steady disks whose viscosity
is given by the ``$\alpha$-model", it is well known that the disk temperature varies as $T \propto M_{\rm bh}^{-1/4}$.
Our models have a steeper dependence, probably due to a different temperature profile resulting from
our viscosity prescription and non-steady conditions.
Model A3 should be compared with model A1, as both have BHs and disrupted stars of the same
mass but differ in  the critical Reynolds number. This affects the instant of maximum temperature and the
late behavior of the temperature decay but not the maximum temperature attained at the peak luminosity.

Snapshots of the spectral evolution of the radiation emitted by the disk derived from model A3 are shown in Figure ~9.
The dashed curve corresponds to the ``true" spectrum, which was calculated taking into account the distribution
of the effective temperature along the disk surface, while the solid curve represents the spectrum
derived from a blackbody distribution characterized by the mean effective temperature as  previously defined.
The upper panel ($t=0.09$ yr) shows the spectrum when the disk is still in formation while the first middle panel
($t=0.60$ yr) shows the spectrum at the luminosity peak. The second middle panel ($t=9.11$ yr) and the bottom panel 
($t=18.23$ yr) show spectra in late evolutionary phases of the disk (or of the flare). Note that as the disk
evolves the mean effective temperature increases, as does the emission at short wavelengths, whereas in the
late stages the opposite behavior is seen. ``Real" and blackbody spectra agree in the Rayleigh-Jeans
region but differ considerably in the Wien's region where the blackbody approximation underestimates
the true flux. In particular, the blackbody approximation underestimates the true X-ray emission from the disk.
As a consequence, models using blackbody distribution to describe the observed
X-ray emission, require temperatures higher than the mean effective values derived from models taking into
account the temperature variation along the disk surface.
We must have this fact in mind when analyzing some flare data in the next section.

\begin{figure}
\plotone{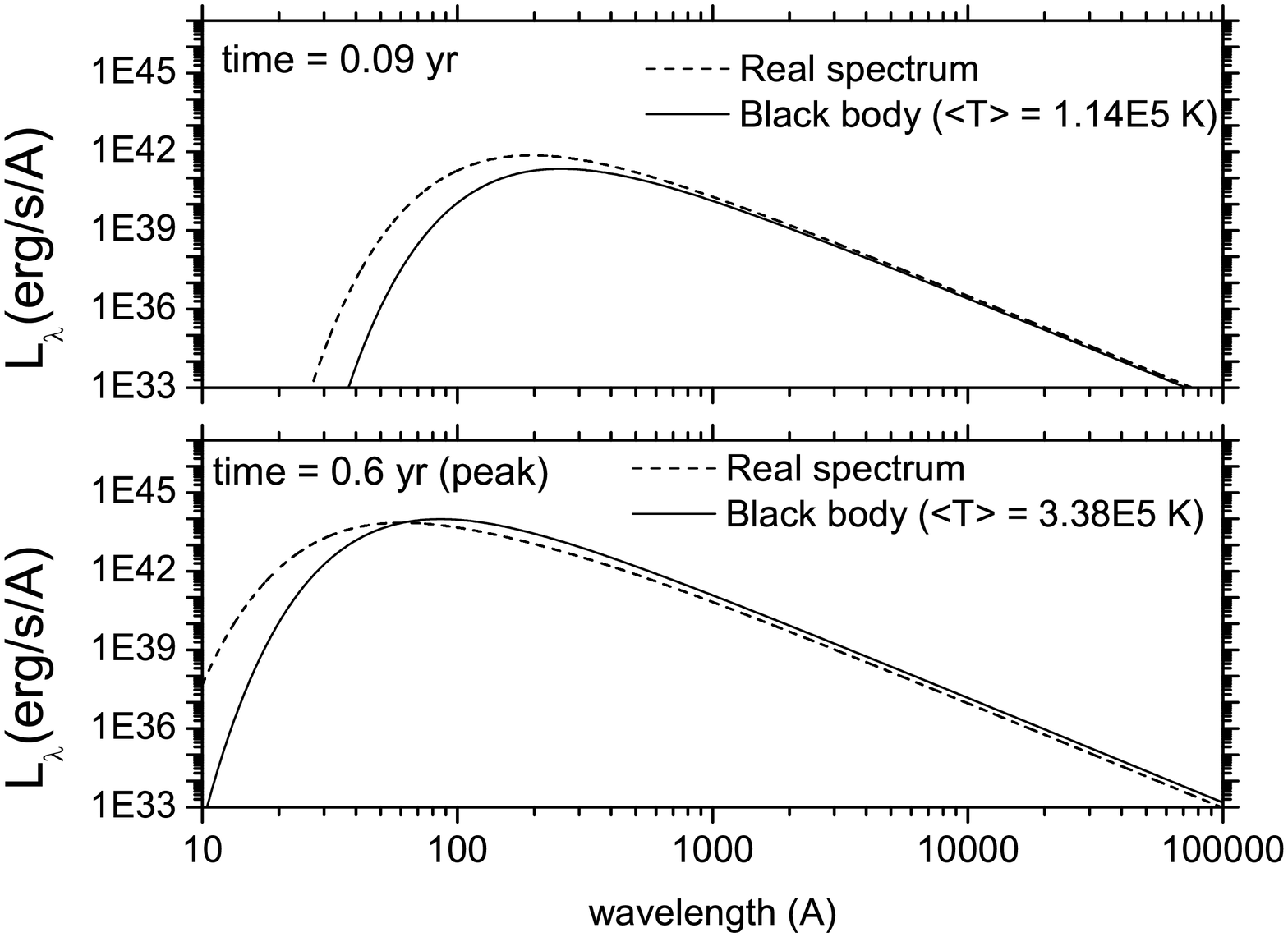}
\vspace{0.1cm}
\plotone{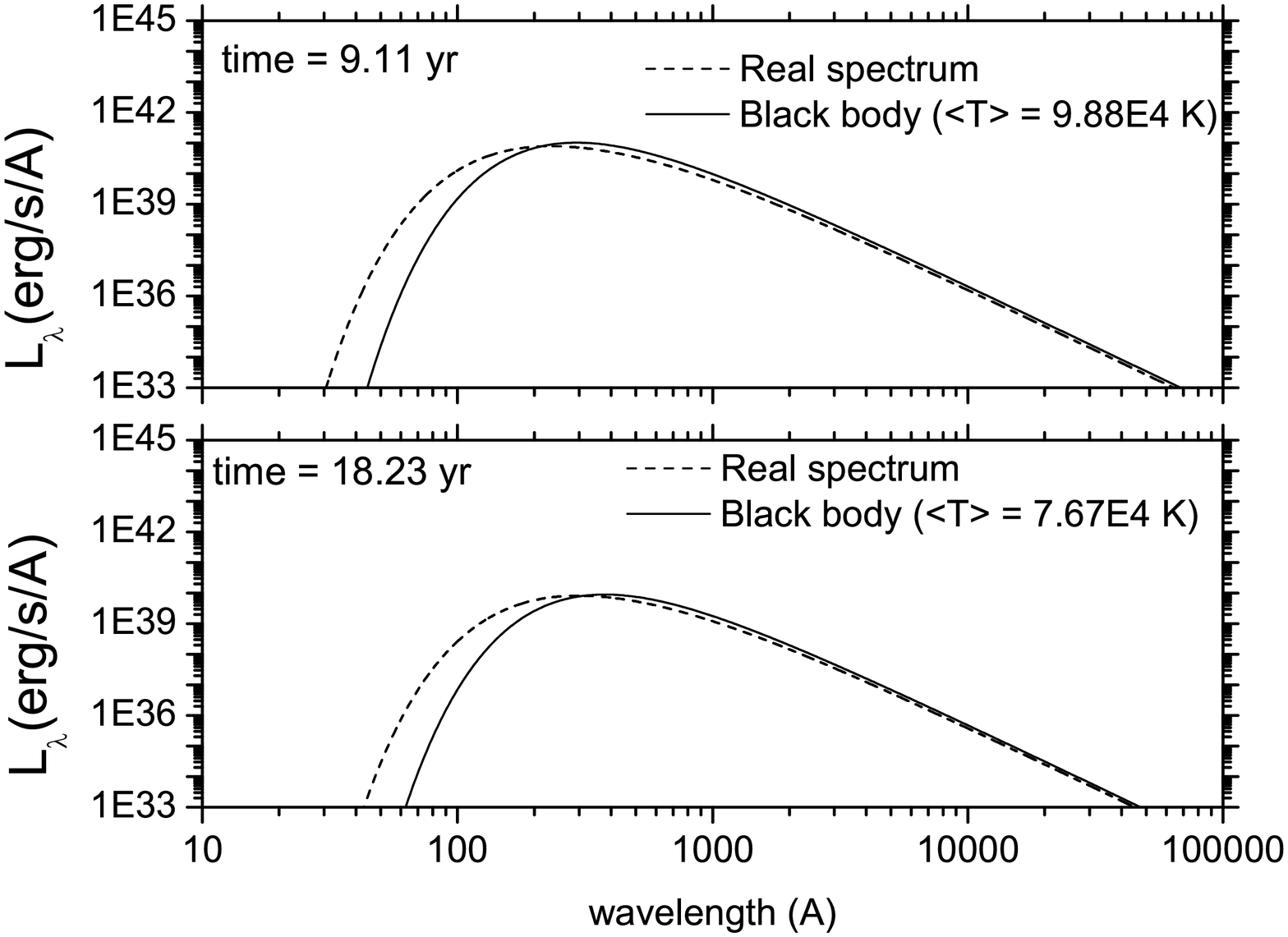}
\caption{Snapshots comparing ``real" (dashed curve) and  equivalent blackbody spectra (solid curve) 
for model A3. The temperature charactering the blackbody spectrum is the average effective temperature calculated 
as explained in the text. The upper panel correspond to a phase in which the disk is still in formation while
the first middle panel shows the spectrum at the luminosity peak. Late evolutionary stages are displayed in the
remaining two panels.}
\end{figure}

\subsection{Comparison with Observations}

Firstly it should be mentioned that the aim of this work was to investigate the evolution of the disk originated
from the debris of a disrupted star by an SMBH and how the different parameters affect the properties of the 
light curve of the resulting outburst. No particular effort was made to fit the observed parameters of different
tidal flare candidates existing in the literature.

\subsubsection{NGC 3599}

NGC 3599 is a non-active S0 galaxy ($z=0.00277$) in which a variable X-ray emission was detected and whose
origin could be due to a tidal disruption event (Esquej et al. 2007, 2008). The source was detected originally
by the \textit{XMM-Newton} slew survey in 2003.89 and was  further observed by this telescope in 2006.49, by
\textit{Swift/X-Ray Telescope} in 2006.94 and by \textit{Chandra} in 2008.1.  Figure ~10 (upper panel)  shows the derived X-ray light curve 
(0.2-2.0 keV range) for model A2 and the X-ray luminosities derived from the aforementioned observations. Note 
that according to our model, the soft X-ray peak luminosity attained a value of $1.8\times 10^{44}~\rm erg ~ s^{-1}$
and occurred about 1.5 yr before detection. 

When the source was first detected, the disk average effective temperature derived from model 
A2, was  is $<kT>=0.016$ keV while the
analysis of \textit{XMM-Newton} data by Esquej et al. (2008) indicates values in the range $0.04 - 0.10$ keV. These
values are higher respectively by factors of two and six than the theoretical average effective temperature but 
such a difference was expected since the disk does not radiate like a blackbody as we have already 
mentioned.

In spite of the fact that the considered model was not optimized, it gives a reasonable description of the
existing data on NGC 3599,  consistent with a tidal disruption event involving a solar mass star
and a BH of mass $\sim 10^7 M_{\odot}$. Note that using the $M_{\rm bh}-\sigma$ relation, the resulting
black hole mass in the center of NGC 3599 is $1.3\times 10^6 M_{\odot}$ and using the $M_{\rm bh}-L_b$ relation
the derived mass is $3.5\times 10^7 M_{\odot}$ (Esquej et al. 2008). 

\begin{figure}
\plotone{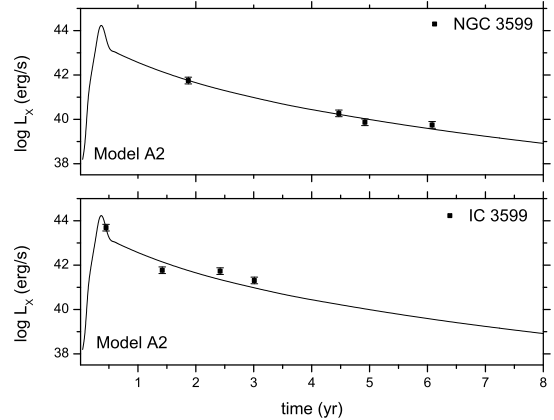}
\caption{Upper panel: evolution of the X-ray luminosity in the 0.2-2.0 keV range for model A2 (solid curve)
compared with data on NGC 3599. The first two observations are from \textit{XMM}, the third from \textit{Swift} and the last from
\textit{Chandra}. Lower panel: evolution of the X-ray luminosity derived also from model A2 and data on IC3599 by \textit{ROSAT}.}
\end{figure}

\subsubsection{IC 3599}

A highly variable X-ray emission originated from the Sb(pec?) galaxy IC 3599 ($z=0.0215$) was detected by the
\textit{ROSAT Wide Field Camera} during its all sky survey (Brandt, Pounds \& Fink 1995). The maximum observed 
luminosity in soft X-rays exceeded $10^{43}~ \rm erg  ~ s^{-1}$ and decayed by two orders of magnitude
in a timescale of about one year. Optical spectra taken 6.3 yr after the X-ray outburst indicate the presence
of strong [OIII] emission and some high-ionization species like HeII$\lambda$4686 and [FeVII]$\lambda$6078,
whose intensities have not changed considerably with respect to data obtained just after the event 
(Komossa \& Bade 1999). The possibility that such an outburst could be associated with a tidal disruption
event was discussed by Grupe et al. (1995) and Komossa \& Bade (1999).  In Figure ~10 (lower panel) we have 
plotted the soft X-ray luminosity curve derived from model A2 and \textit{ROSAT} data for IC 3599 
taken from Grupe et al. (1995). Again, we emphasize
that the considered model gives a good representation of these data but parameters were not optimized in order
to obtain a ``best fit". If correct, the model indicates an SMBH in the center of IC 3599 with a mass around
$10^7~M_{\odot}$ and that the disrupted star had a mass close to the solar value. Comparison between 
the theoretical X-ray light curve and data also indicates  that detection occurred close to the luminosity 
peak ($\sim$ 20 days after) and about 5 months after the beginning 
of the circularization process. The mean effective temperatures derived from model A2
corresponding to the considered data points are respectively 26.6 eV, 15.9 eV, 15.8 eV and 14.4 eV. These  should
be compared with the values given by Grupe et al. (1995), i.e., 94 eV, 65 eV, 80 eV and 62 eV. These values
are, on the average, a factor of four higher than those derived from model A2, confirming the trend already discussed.

\begin{figure}
\plotone{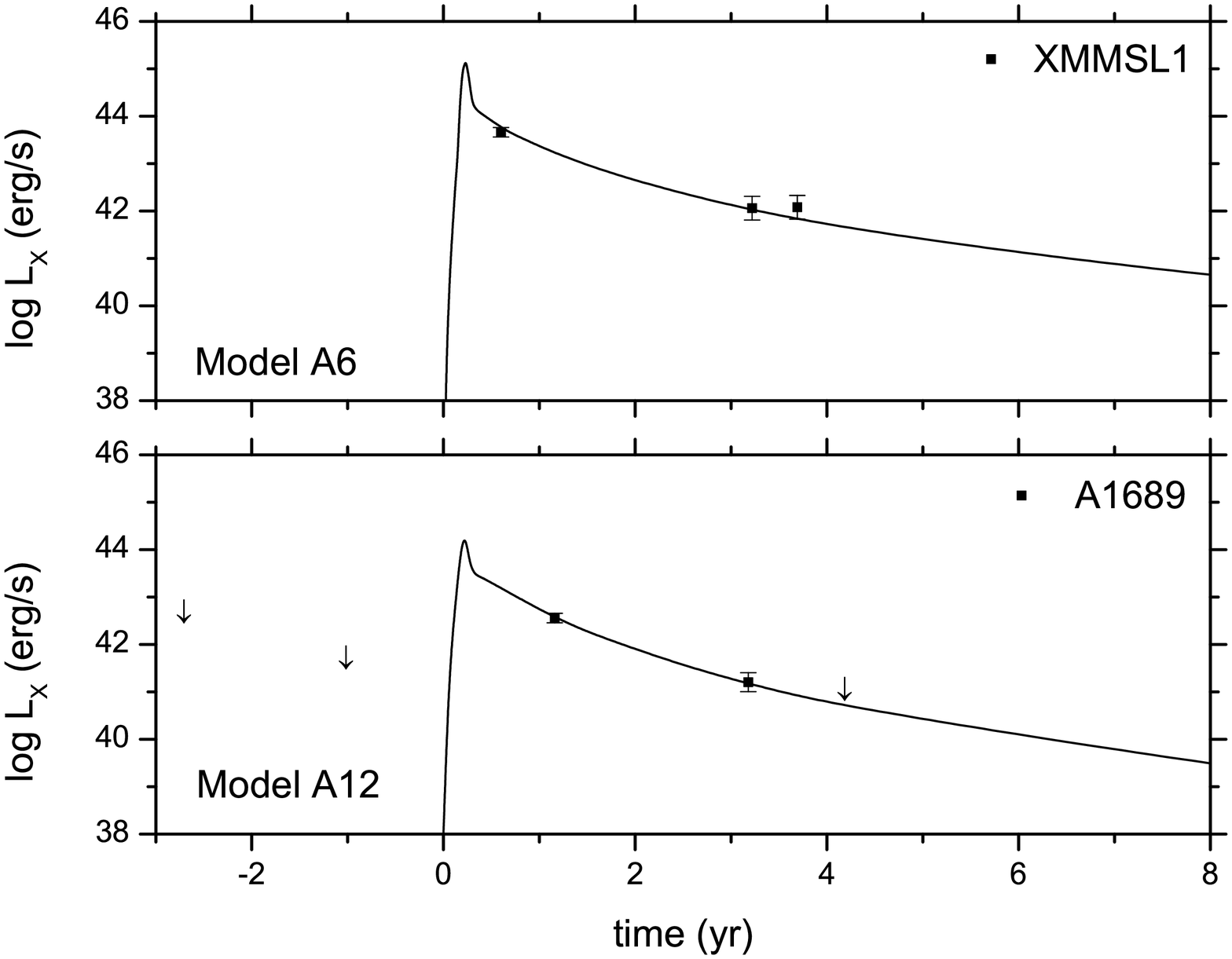}
\caption{Upper panel: evolution of X-ray luminosity in the 0.2-2.0 keV range for model A6 (solid curve)
compared with data on XMMSL1. The first two observations are from \textit{XMM-Newton} and the third from \textit{Swift}. 
Lower panel: evolution of the X-ray luminosity derived from model A12 and data on the outburst
detected in the cluster A1689 by \textit{Chandra} and \textit{XMM-Newton}. Small arrows indicate upper limits.}
\end{figure}

\subsubsection{XMMSL1}

The X-ray emission of XMMSL1 (or SDSS J132342.3 +482701) was detected in 2003.92 by \textit{XMM-Newton} (Esquej et al.
2008). The variable X-ray emission is originates from a non-active galaxy whose spectrum displays only
absorption features. This galaxy is at redshift $z=0.0875$, and its stellar velocity dispersion is
$80\pm 10~\rm km ~ s^{-1}$ (Esquej et al. 2008). Using the $M_{\rm bh}-\sigma$ relation, the measured velocity
dispersion implies that the central black hole has a mass around $2\times 10^6~M_{\odot}$. Model
A6 is defined by a black hole of similar mass ($3\times 10^6~M_{\odot}$) and a disrupted star of one
solar mass. In Figure ~11 (upper panel), the X-ray luminosity curve derived from this model is shown in
comparison with data from \textit{XMM-Newton} and \textit{Swift}. The model is consistent with observations, but the
small  number of data points avoids a more robust conclusion. If correct, the model indicates that
detection occurred 5-6 months after the X-ray peak luminosity.

X-ray data at the high-state emission if fitted by a blackbody spectrum indicate temperatures in the
range 33-49 eV (Esquej et al. 2008) while we have derived from model A6 an average effective temperature
of 30 eV.

\subsubsection{A1689}

\textit{Chandra} observations in 2004.14 detected a variable X-ray emission associated with the 
galaxy SDSS J131122.15 -012345.6, a member of the cluster A1689 at a redshift of 
$z=0.183$ (Maksym, Ulmer \& Eracleous 2010). Prior observations performed either in 2000.29 and in 2001.98 by 
\textit{Chandra} and \textit{XMM-Newton} have not detected the source. Since only two data points and three upper
limits are available, it is difficult to constrain a model. The X-ray light curve derived from model
A12, characterized by a $10^7~M_{\odot}$ and a disrupted star of half solar mass gives an acceptable
representation of the existing data, satisfying also the pre-outburst constraints (see Figure ~11, lower
panel).

In its high-state emission Maksym, Ulmer \& Eracleaous -2010) derived from their fits an equivalent
blackbody temperature of 120 eV whereas the average mean temperature derived from model A12 is 21.6 eV.
This  is again consistent with our previous statement that blackbody fits to X-ray data overestimates the
true mean temperature of the disk.

\subsubsection{NGC 5905}

NGC 5905 is a SB(r)b galaxy located at $z=0.0113$. A variable X-ray emission from this object
was detected in the all sky survey by \textit{ROSAT} around 1990 July  (Bade, Komossa \& Dahlem 1996). 
Subsequent observations in the following six days indicated an increase in the flux by a factor
of three and pointed observations performed in 1993 showed that the X-ray emission decayed by
almost two orders of magnitude, suggesting the occurrence of an outburst event.
Its optical spectrum is dominated by strong absorption lines, and narrow forbidden lines ([OI], [OIII], [NII], 
and [SII]), as well as Balmer emission lines (Komossa \& Bade 1999). Since no particular signals of nuclear 
activity are seen, the galaxy may have a ``dormant" central SMBH that could be responsible for a tidal 
disruption event associated with the X-ray outburst detected by \textit{ROSAT}. 

In order to analyze the existing X-ray data on this event, we recall that the rising part of the light
curve in our models is determined by the accretion (or viscous) timescale, which is typically of the
order of 4 months. If the observed increase in the X-ray flux seen in the early detection by \textit{ROSAT} corresponds to
the raising part of the flare, implying a timescale of $\sim$ 5 days, then none of our models are able to fit 
the X-ray light curve. Such a short raising timescale implies a very small disk that could be formed if the disrupted
object was a low-mass star. For instance, a M8V star with a mass of $\sim 0.1 M_{\odot}$ and a radius of
$\sim 0.12 R_{\odot}$ will correspond, using Equation ~\ref{fitpeak}, to an accretion timescale of about 10 days for a 
critical Reynolds number equal to 500. The analysis by Li, Narayan \& Menou (2002) leads to a similar conclusion. 

\section{Conclusions}
 
In the present paper we report the results derived from numerical solutions of the hydrodynamic equations
describing the evolution of an accretion disk originated from the debris resulting from a tidal disruption
event. Models are characterized by the central black hole mass, the properties of the disrupted star
(mass and radius) and the critical Reynolds number defining the onset of a turbulent flow.

Once the bound material begins to circularize, there is a finite amount of time before the black hole begins to
accrete mass, which is defined by the accretion (or ``viscous") timescale $t_a$. In the beginning of the 
circularization process a torus-like structure is formed close to the periapse level (assumed here to coincide
with the tidal radius), expanding inward until the last stable orbit is reached in a timescale $\sim t_a$.
This initial phase, when the disk is still being formed, corresponds to the rising part of the light curve
that characterizes the event. The peak luminosity marks the beginning of the accretion process by the
central black hole,  consequently also occurring  in a timescale $t_a$. The presence or lack of a ``bump"
in the light curve after the luminosity peak depends on the ratio $t_a/t_{\rm max}$ or, in other words, how the accretion
timescale compares with the instant $t_{\rm max}$ where the maximum of the fallback rate occurs. ``Bumps"
occur when $t_a/t_{\rm max} < 1$, corresponding to models of series ``M". They are absent in models of series ``A",
satisfying the opposite condition. The comparison with actual observations leads to the conclusion that ``A"-models
are more adequate to represent the existing data. 

Peak bolometric luminosities for ``A"-models are in the range $10^{45}-10^{46}~ \rm erg ~ s^{-1}$
while peak luminosities in soft X-rays (0.2-2.0 keV) are one order of magnitude lower. For the same parameters,
models of series ``M" have peak luminosities about one order of magnitude lower. The characteristic
timescale for the luminosity to decay by two orders of magnitude after the peak is about 3.3 yr for
models ``A" and 43.3 yr for models ``M". These values are considerably smaller than the timescales derived
by Cannizzo, Lee \& Goodman (1990) from their models.
  
The effective temperature varies along the disk surface as well as in time. The maximum values are observed
when the debris material reaches the last stable orbit for the first time, which is of the order of $(1-2)\times
10^6$K. At this particular moment, the most important temperature gradient across the disk surface is
observed. In the last evolutionary phases, the temperature distribution along the disk surface becomes
more uniform and the disk becomes more and more ``cold" as the source of debris is exhausted.

The resulting spectrum is not that of a blackbody since it results from the integrated emission
from a surface having a temperature distribution and, despite the fact that the effective
optical depth is much larger than unity, scattering by free electrons and free-free absorption
contributes equally to the opacity and, consequently, the local spectrum differs from a blackbody.
Comparison between true spectra and those computed with a blackbody characterized by an average temperature
giving the same total disk emission, indicates that they agree in the Rayleigh-Jeans region
but that  the blackbody underestimates the true flux at short wavelengths. Consequently, blackbody fits
of X-ray data tend to overestimate the true mean disk temperature. In fact, the temperature derived
from the blackbody fits should be identified with the color X-ray temperature rather than the true
average disk effective temperature.

In spite of the fact that the present models were not computed to explain any particular observation,
they were tested for several tidal flare event candidates. X-ray data on NGC 3599 and IC 3599 can
be satisfactorily represented by  model A2, which  involves a black hole of $10^7~M_{\odot}$ and a
disrupted star of one solar mass. According to our model, NGC 3599 was first detected $\sim 1.5$ yr
after the X-ray luminosity peak and IC 3599 very close to it ($\sim 20$ days after). The soft
X-ray light curve derived from Model A6, characterized by a $3\times 10^6~M_{\odot}$ black hole and a 
disrupted star of one solar mass  fits quite well data on XMMSL1. The X-ray peak emission for this object occurred
probably 5-6 months before detection at a luminosity of about $10^{45}~ \rm erg  ~ s^{-1}$. The outburst that
occurred in the cluster A1689 is well represented by model A12 involving a $10^7~M_{\odot}$ black hole
but the disrupted star is a low-mass object with half solar mass. Its detection occurred about one year
after the peak luminosity according to our model. Finally, none of our models were able to fit the existing
data on NGC 5905. The short rise time of the flare points to the disruption and capture of the debris
of a low-mass object, and  this problem is presently under investigation.

\subsection{Acknowledgments}

M. M.A acknowledges the Comisi\'on Nacional de Investigaci\'on Cientifica y Tecnol\'ogica (CONICYT)
de Chile for the fellowship that permitted his stay at the Observatoire de la C\^ote d'Azur.
The authors thank the referee for his/her  suggestions that have improved the presentation of this
paper.



\end{document}